\documentclass[aps,prd,preprint,tightenlines,nofootinbib,showpacs]{revtex4}

\usepackage{epsfig}
\usepackage{graphicx}
\usepackage{graphics}
\usepackage{dcolumn}
\usepackage{bm}
\usepackage{afterpage}

\newcommand{\etal}{{\it et al.}}

\begin{document}

\preprint{\tighten\vbox{\hbox{\hfil CLNS 07/1990}
                        \hbox{\hfil CLEO 07-02}
}}

\title{Measurement of  ${\cal{B}}(D_s^+\to\ell^+\nu)$ and the Decay Constant
$f_{D_s^+}$\hspace*{2mm}}

\author{T.~K.~Pedlar}
\affiliation{Luther College, Decorah, Iowa 52101}
\author{D.~Cronin-Hennessy}
\author{K.~Y.~Gao}
\author{J.~Hietala}
\author{Y.~Kubota}
\author{T.~Klein}
\author{B.~W.~Lang}
\author{R.~Poling}
\author{A.~W.~Scott}
\author{A.~Smith}
\author{P.~Zweber}
\affiliation{University of Minnesota, Minneapolis, Minnesota 55455}
% ...
\author{(CLEO Collaboration)} 
\collaboration{CLEO Collaboration} 
\noaffiliation

\date{April 2, 2003}

\begin{abstract}
We examine $e^+e^-\to D_s^-D_s^{*+}$ and $D_s^{*-}D_s^{+}$
interactions at 4170 MeV using the CLEO-c detector in order to
measure the decay constant $f_{D_s^+}$. We use the $D_s^+\to
\ell^+\nu$ channel, where the $\ell^+$ designates either a $\mu^+$
or a $\tau^+$, when the $\tau^+\to\pi^+\overline{\nu}$. Analyzing
both modes independently, we determine ${\cal{B}}(D_s^+\to
\mu^+\nu)= (0.594\pm 0.066\pm0.031)$\%, and ${\cal{B}}(D_s^+\to
\tau^+\nu)= (8.0\pm 1.3\pm0.4)$\%. We also analyze them
simultaneously to find an effective value of
${\cal{B}}^{eff}(D_s^+\to \mu^+\nu)= (0.638 \pm 0.059 \pm 0.033)$\%
and extract $f_{D_s^+}=274\pm 13 \pm 7 {~\rm MeV}$. Combining with
our previous determination of ${\cal{B}}(D^+\to \mu^+\nu)$, we also
find the ratio $f_{D_s^+}/f_{D^+}=1.23\pm 0.11\pm 0.04$. We compare
to current theoretical estimates. Finally, we find
${\cal{B}}(D_s^+\to e^+\nu) < 1.3\times 10^{-4}$ at 90\% confidence
level.
\end{abstract}

\pacs{13.20.Fc, 13.66.Bc}
\maketitle \tighten


\section{Introduction}

To extract precise information on the size of
Cabibbo-Kobayashi-Maskawa matrix elements from $B-\overline{B}$
mixing measurements, the ``decay constants" for $B_d$ and $B_s$
mesons or their ratio, $f_{B_d}/f_{B_s}$, must be well known
\cite{formula-mix}. These factors are related to the overlap of the
heavy and light quark wave-functions at zero spatial separation.
Indeed, the recent measurement of $B_s^0-\overline{B}_s^0$ mixing by
CDF \cite{CDF} that can now be compared to the very well measured
$B^0$ mixing \cite{PDG} has pointed out the urgent need for precise
values \cite{Belle-taunu}. Decay constants have been calculated
theoretically. The most promising of these calculations are based on
lattice-gauge theory that includes light quark loops \cite{Davies},
often called ``unquenched."  In order to ensure that these theories
can adequately predict $f_{B_s}/f_{B_d}$ it is useful to check the
analogous ratio in charm decays $f_{D^+_s}/f_{D^+}$. We have
previously measured $f_{D^+}$ \cite{our-fDp,DptomunPRD}. Here we
present the most precise measurement to date of $f_{D_s^+}$ and the
ratio $f_{D_s^+}/f_{D^+}$.

\begin{figure}[htbp]
 \vskip 0.00cm
 \centerline{ \epsfxsize=3.0in \epsffile{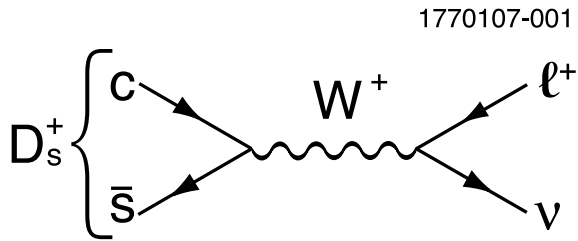} }
 \caption{The decay diagram for $D_s^+\to \ell^+\nu$.} \label{Dstomunu}
 \end{figure}

In the Standard Model (SM), the only way for a $D_s$ meson to decay
purely leptonically is via annihilation through a virtual $W^+$, as
depicted in Fig.~\ref{Dstomunu}. The decay rate is given by
\cite{Formula1}
\begin{equation}
\Gamma(D_s^+\to \ell^+\nu) = {{G_F^2}\over
8\pi}f_{D_s^+}^2m_{\ell}^2M_{D_s^+} \left(1-{m_{\ell}^2\over
M_{D_s^+}^2}\right)^2 \left|V_{cs}\right|^2~~~, \label{eq:equ_rate}
\end{equation}
where $M_{D_s^+}$ is the $D_s^+$ mass, $m_{\ell}$ is the mass of the
charged final state lepton, $G_F$ is the Fermi coupling constant,
and $|V_{cs}|$ is a Cabibbo-Kobayashi-Maskawa matrix element with a
value we take equal to 0.9738 \cite{PDG}.

In this paper we analyze both $D_s^+\to\mu^+\nu$ and
$D_s^+\to\tau^+\nu$, $\tau^+\to \pi^+\overline{\nu}$. In both $D_s$
decays the charged lepton must be produced with the wrong helicity
because the $D_s$ is a spin-0 particle, and the final state consists
of a naturally left-handed spin-1/2 neutrino and a naturally
right-handed spin-1/2 anti-lepton. Because the $\tau^+$ has a mass
close to that of the $D_s^+$, the helicity suppression is broken
with respect to the $\mu^+$ decay, but there is an additional large
phase space suppression.

New physics can affect the expected widths; any undiscovered
charged bosons would interfere with the SM $W^+$. These effects
may be difficult to ascertain, since they would simply change the
value of the decay constants. The ratio $f_{D_s^+}/f_{D^+}$ is
much better predicted in the SM than the values individually, so
deviations from the the SM expectation are more easily seen. Any
such discrepancies would point to beyond the SM charged bosons.
For example, Akeroyd predicts that the presence of a charged Higgs
boson would suppress this ratio significantly \cite{Akeroyd}.

We can also measure the ratio of decay rates to different leptons,
and the SM predictions then are fixed only by well-known masses. For
example, for $\tau^+\nu$ to $\mu^+\nu$:

\begin{equation}
R\equiv \frac{\Gamma(D_s^+\to \tau^+\nu)}{\Gamma(D_s^+\to
\mu^+\nu)}= {{m_{\tau^+}^2 \left(1-{m_{\tau^+}^2\over
M_{D_s^+}^2}\right)^2}\over{m_{\mu^+}^2 \left(1-{m_{\mu^+}^2\over
M_{D_s^+}^2}\right)^2}}~~. \label{eq:tntomu}
\end{equation}
Using measured masses \cite{PDG}, this expression yields a value of
9.72 with a negligibly small error.
Any deviation in $R$ from the value predicted by Eq.~\ref{eq:tntomu}
would be a manifestation of physics beyond the SM. This could occur
if any other charged intermediate boson existed that affected the
decay rate differently than mass-squared. Then the couplings would
be different for muons and $\tau$'s. This would be a clear violation
of lepton universality \cite{Hewett}.

Previous measurements of $f_{D_s^+}$ have been hampered by a lack of
statistical precision, and relatively large systematic errors
\cite{CLEO,BEAT,ALEPH,L3,OPAL,Babar-munu}. One large systematic
error source has been the lack of knowledge of the absolute
branching fraction of the normalization channel, usually
$D_s^+\to\phi\pi^+$ \cite{stone-fpcp}. The results we report here
will not have this limitation.

\section{Experimental Method}
\subsection{Selection of $D_s$ Candidates}
The CLEO-c detector \cite{CLEODR} is equipped to measure the momenta
and directions of charged particles, identify them using specific
ionization ($dE/dx$) and Cherenkov light (RICH) \cite{RICH}, detect
photons and determine their directions and energies.

In this study we use 314 pb$^{-1}$ of data produced in $e^+e^-$
collisions using the Cornell Electron Storage Ring (CESR) and
recorded near a center-of-mass energy ($E_{\rm CM}$) of 4.170 GeV.
At this energy the $e^+e^-$ annihilation cross-section into
$D_s^{*+}D_s^-$+$D_s^{+}D_s^{*-}$ is approximately 1~nb, while the
cross-section  for $D_s^+D_s^-$ is about a factor of 20 smaller. In
addition, $D$ mesons are produced mostly as $D^{*}\overline{D^*}$,
with a cross-section of $\sim$5~nb, and also in
$D^*\overline{D}+D\overline{D^*}$ final states with a cross-section
of $\sim$2 nb. The $D\overline{D}$ cross-section is a relatively
small $\sim$0.2 nb \cite{poling}. There also appears to be
$D\overline{D}^*\pi$ production. The underlying light quark
``continuum" background is about 12 nb. The relatively large
cross-sections, relatively large branching fractions and sufficient
luminosities allow us to fully reconstruct one $D_s$ as a ``tag,"
and examine the properties of the other. In this paper we designate
the tag as a $D_s^-$ and examine the leptonic decays of the $D_s^+$,
though, in reality, we use both charges for tags and signals. Track
requirements, particle identification, $\pi^0$, $\eta$, and $K_S^0$
selection criteria are the same as those described in
Ref.~\cite{our-fDp}, except that we now require a minimum momentum
of 700 MeV/c for a track to be identified using the RICH.

We also use several resonances that decay via the strong
interaction.
Here we select intervals in invariant mass within $\pm$10 MeV of the
known mass for $\eta'\to\pi^+\pi^-\eta$, $\pm$10 MeV for $\phi\to
K^+ K^-$, $\pm$100 MeV for $K^{*0}\to K^-\pi^+$, and $\pm$150 MeV
for $\rho^-\to \pi^-\pi^0$.

\begin{figure}[htb]
\includegraphics[width=74mm]{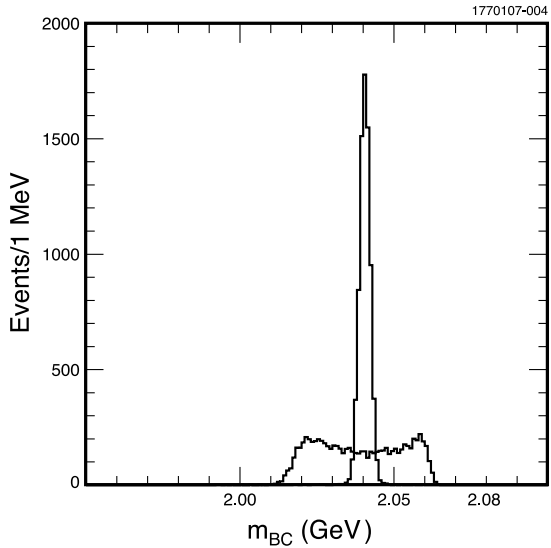}
\vspace{0.44mm}\caption{The beam constrained mass $m_{\rm BC}$ from
Monte Carlo simulation of $e^+e^-\to D_s^+D_s^{*-}$,
$D_s^{\pm}\to\phi\pi^{\pm}$ at an $E_{\rm CM}$ of 4170 MeV. The
narrow peak is from the $D_s^+$ and the wider one from
$D_s^{*-}\to\gamma D_s^-$. (The distributions are not centered at
the $D_s^+$ or $D_s^{*+}$ masses, because the reconstructed
particles are assumed to have the energy of the beam.)} \label{mbc}
\end{figure}

We reconstruct tags from either directly produced $D_s$ mesons or
those that result from the decay of a $D_s^*$. The beam constrained
mass, $m_{\rm BC}$, is formed by using the beam energy to construct
the $D_s$ candidate mass via the formula
\begin{equation}
m_{\rm BC}=\sqrt{E_{\rm beam}^2-(\sum_i\overrightarrow{p}_{\!i})^2},
\end{equation}
where $i$ runs over all the final state particles. If we ignore the
photon from the $D_s^*\to\gamma D_s$ decay, and reconstruct the
$m_{\rm BC}$ distribution, we obtain the distribution from Monte
Carlo simulation shown in Fig.~\ref{mbc}. The narrow peak occurs
when the reconstructed $D_s$ does not come from the $D_s^*$ decay,
but is directly produced.

\begin{figure}[hbtp]
\centering
\includegraphics[width=5in]{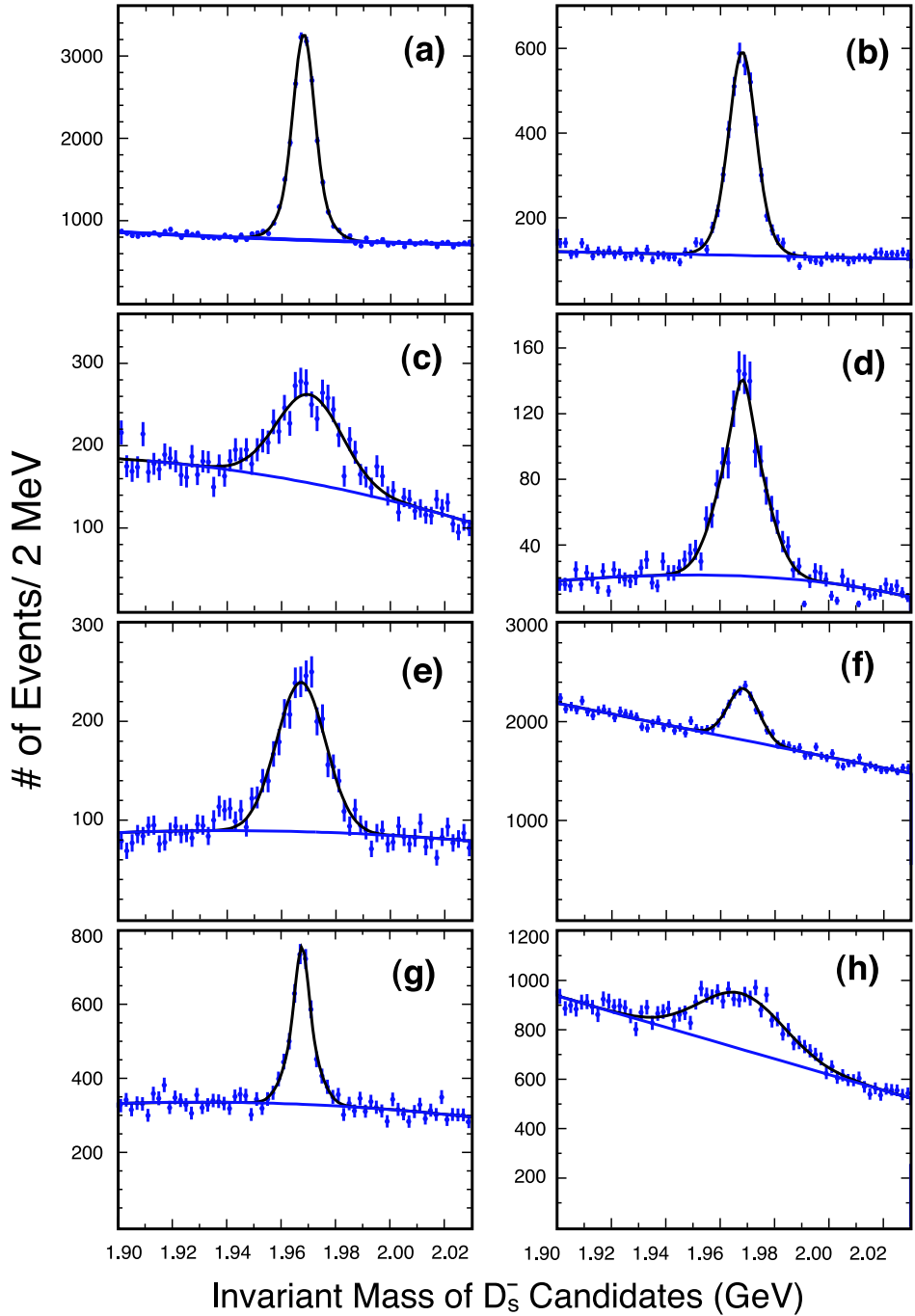}
\caption{Invariant mass of $D_s^-$ candidates in the decay modes (a)
$K^+K^-\pi^-$, (b) $K_SK^-$, (c) $\eta\pi^-$, (d) $\eta'\pi^-$, (e)
$\phi\rho^-$, (f) $\pi^+\pi^-\pi^-$, (g) $K^{*-}K^{*0}$, and (h)
$\eta\rho^-$, after requiring the total energy of the $D_s^-$
candidate to be consistent with the beam energy. The curves are fits
to two-Gaussian signal functions plus a polynomial background.
 } \label{Inv-mass}
\end{figure}

Rather than selecting events based on only $m_{\rm BC}$, we first
select an interval that accepts most of the events, $2.015 <m_{\rm
BC}<2.067$ GeV, and examine the invariant mass. Distributions from
data for the 8 tag decay modes we use in this analysis are shown in
Fig.~\ref{Inv-mass}. Note that the resolution in invariant mass is
excellent, and the backgrounds not abysmally large, at least in
these modes. To determine the number of $D_s^-$ events we fit the
invariant mass distributions to the sum of two Gaussians centered at
the $D_s^-$ mass, a function we refer to as ``two-Gaussian." The
r.m.s. resolution ($\sigma$) is defined as
\begin{equation}
\sigma \equiv f_1\sigma_1+(1-f_1)\sigma_2, \label{eq:twoGauss}
\end{equation}
where $\sigma_1$ and $\sigma_2$ are the individual widths of each
of the two Gaussians and $f_1$ is the fractional area of the first
Gaussian. The number of tags in each mode is listed in
Table~\ref{tab:Ntags}. We will later use sidebands of the signal
peaks shown in Fig.~\ref{Inv-mass} for part of the background
estimate.

\begin{table}[htb]
\begin{center}
\caption{Tagging modes and numbers of signal and background events,
within $\pm2.5\sigma$ of the $D_s^-$ mass for all modes, except
$\eta\rho^-$ ($\pm 2\sigma$), determined from two-Gaussian fits to
the invariant mass plots, and the number of photon tags in each
mode, within $\pm2.5\sigma$ of the $D_s$ mass-squared determined
from fits of the MM$^{*2}$ distributions (see text) to a signal
Crystal Ball function (see text) and a 5th order Chebychev
background polynomial, and the associated
backgrounds.\label{tab:Ntags}}
\begin{tabular}{lcccc}
 \hline\hline
    Mode  & \multicolumn{2}{c}{Invariant Mass}& \multicolumn{2}{c}{MM$^{*2}$}\\
    &  Signal & Background & Signal & Background \\\hline
$K^+K^-\pi^- $ & 13871$\pm$262 & 10850 & 8053$\pm$ 211 &13538\\
$K_S K^-$ & 3122$\pm$79 & 1609 & 1933$\pm$88&2224\\
$\eta\pi^-$; $\eta\to\gamma\gamma$ & $1609\pm 112$  &
4666&1024$\pm$97 &3967\\
$\eta'\pi^-$; $\eta'\to\pi^+\pi^-\eta$, $\eta\to\gamma\gamma$ & 1196$ \pm $46  &409 &792$\pm$69 &1052 \\
$\phi\rho^-$; $\phi\to K^+K^-$, $\rho^-\to \pi^-\pi^0$ & 1678$ \pm $74  &1898&1050$\pm$113&3991 \\
$\pi^+\pi^-\pi^-$ & 3654$ \pm $199  & 25208 & 2300$\pm$187& 15723\\
$K^{*-}K^{*0}$; $K^{*-}\to K_S^0\pi^-$, ${K}^{*0}\to K^+\pi^-$ &
2030$
\pm$98& 4878&1298$\pm$130 & 5672\\
$\eta\rho^-$; $\eta\to\gamma\gamma$, $\rho^-\to \pi^-\pi^0$ & 4142$ \pm $281  &20784 & 2195$\pm$225 & 17353\\
\hline
Sum &  $31302\pm 472 $ &70302 & 18645$\pm$426&63520\\
\hline\hline
\end{tabular}
\end{center}
\end{table}

To select our sample of tag events, we require the invariant
masses, shown in Fig.~\ref{Inv-mass}, to be within $\pm
~2.5\sigma$ ($\pm2\sigma$ for the $\eta\rho^-$ mode) of the known
$D_s^-$ mass. Then we look for an additional photon candidate in
the event that satisfies our shower shape requirement. Regardless
of whether or not the photon forms a $D_s^*$ with the tag, for
real $D_s^*D_s$ events, the missing mass squared, MM$^{*2}$,
recoiling against the photon and the $D_s^-$ tag should peak at
the $D_s^{+}$ mass-squared. We calculate
\begin{equation}
\label{eq:mmss} {\rm MM}^{*2}=\left(E_{\rm
CM}-E_{D_s}-E_{\gamma}\right)^2- \left(\overrightarrow{p}_{\!\rm
CM}-\overrightarrow{p}_{\!D_s}-\overrightarrow{p}_{\!\gamma}\right)^2,
\end{equation}
where $E_{\rm CM}$ ($\overrightarrow{p}_{\!\rm CM}$) is the
center-of-mass energy (momentum), $E_{D_s}$
($\overrightarrow{p}_{\!D_s}$) is the energy (momentum) of the fully
reconstructed $D_s^-$ tag, and $E_{\gamma}$
($\overrightarrow{p}_{\!\gamma}$) is the energy (momentum) of the
additional photon. In performing this calculation we use a kinematic
fit that constrains the decay products of the $D_s^-$ to the known
$D_s$ mass and conserves overall momentum and energy. All photon
candidates in the event are used, except for those that are decay
products of the $D_s^-$ tag candidate.

\begin{figure}[hbt]
\centering
\includegraphics[width=4.4in]{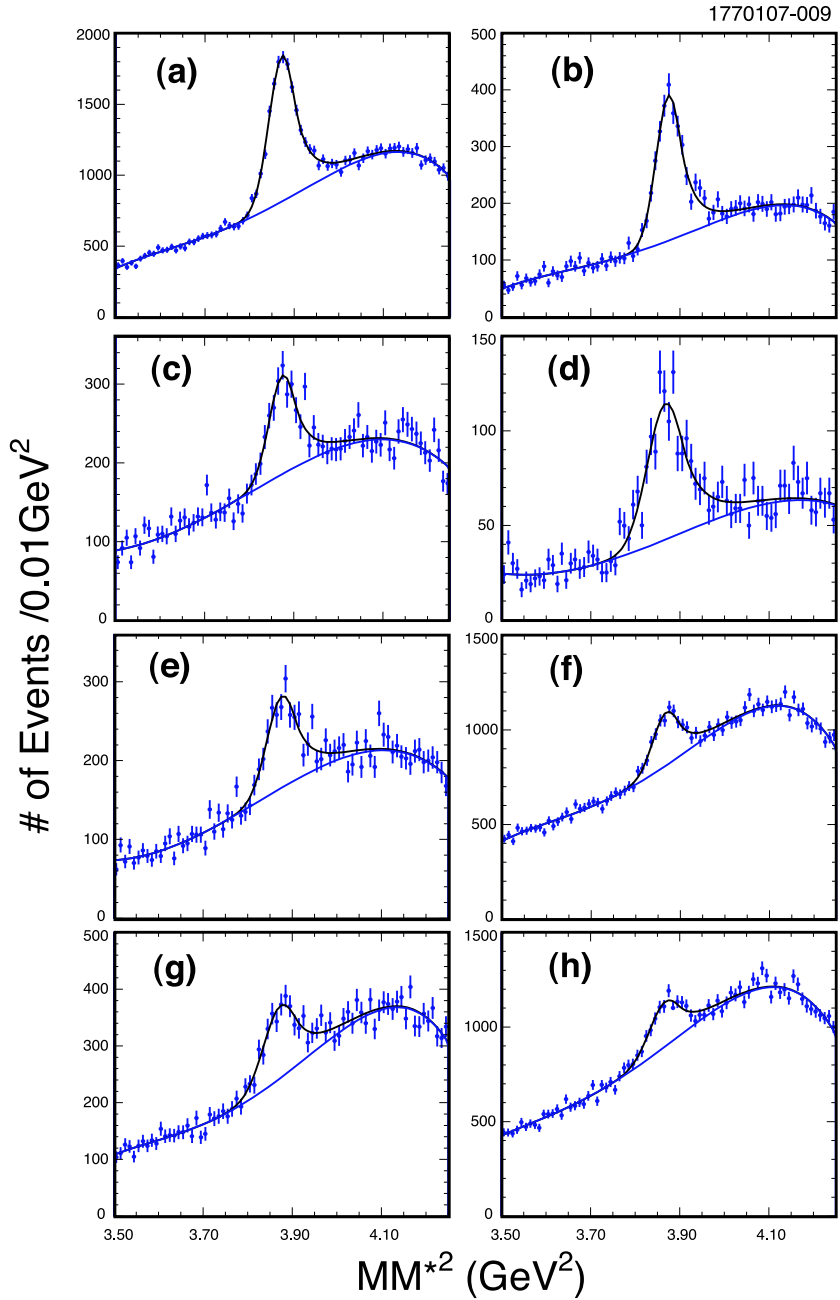}
\caption{The MM$^{*2}$ distribution from events with a photon in
addition to the $D_s^-$ tag for the modes: (a) $K^+K^-\pi^-$, (b)
$K_S^0K^-$, (c) $\eta\pi^-$, (d) $\eta'\pi^-$, (e) $\phi\rho^-$, (f)
$\pi^+\pi^-\pi^-$, (g) $K^{*-}K^{*0}$, and (h) $\eta\rho^-$. The
curves are fits to the Crystal Ball function and a 5th order
Chebychev background function.} \label{MMstar2}
\end{figure}

\begin{figure}[hbt]
\centering
\includegraphics[width=4in]{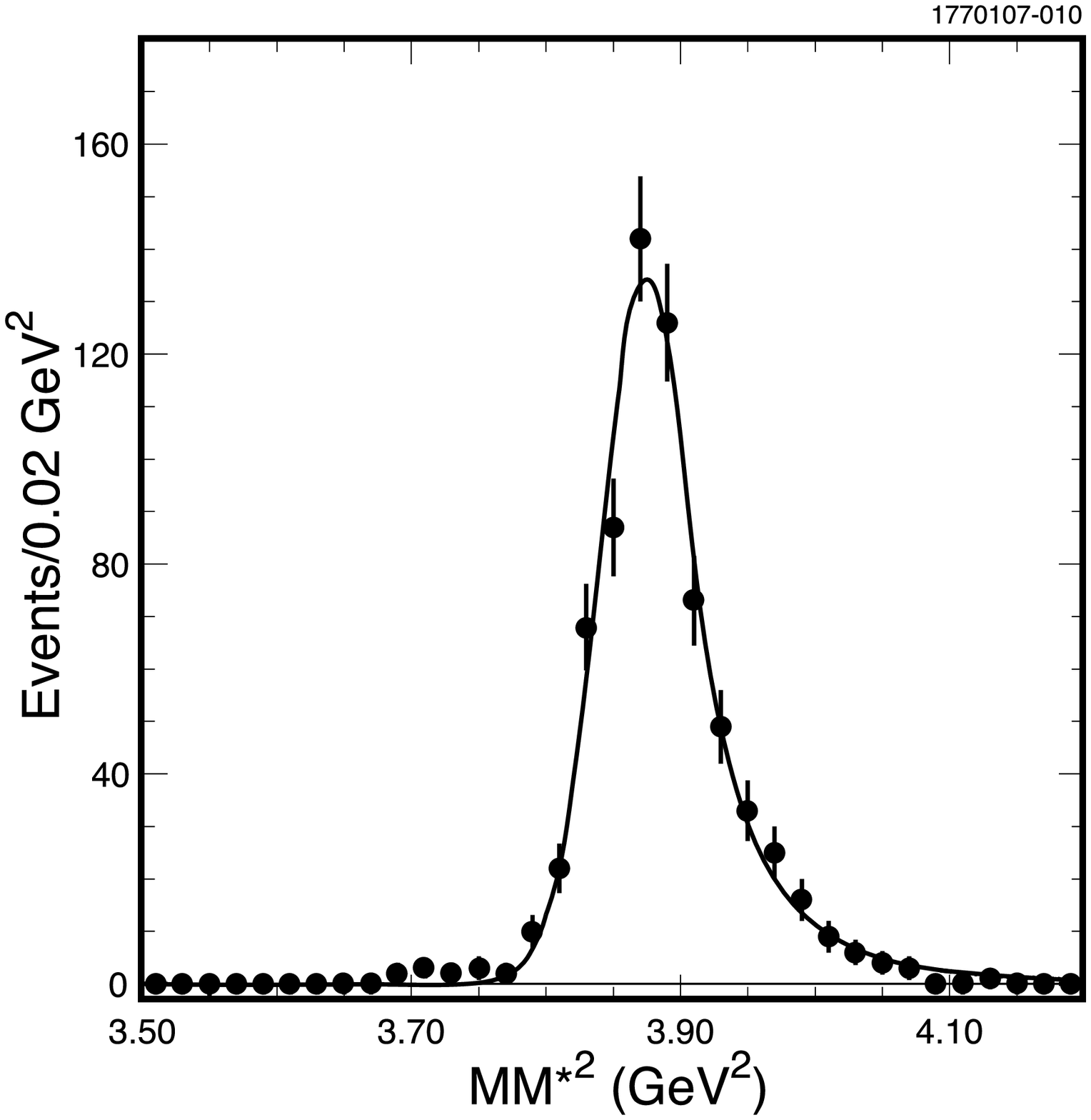}
\caption{The MM$^{*2}$ distribution from a sample of fully
reconstructed $D_s^- D_s^{*+}$ events where one $D_s$ is ignored.
The curve is a fit to the Crystal Ball function.}
\label{data-Ds-MM2-DoubleTags}
\end{figure}

The MM$^{*2}$ distributions from the selected $D_s^-$ event sample
are shown in Fig.~\ref{MMstar2}. We fit these distributions to
determine the number of tag events. This procedure is enhanced by
having information on the shape of the signal function. One
possibility is to use the Monte Carlo simulation for this purpose,
but that would introduce a relatively large systematic error.
Instead, we use our relatively large sample of fully reconstructed
$D_s D_s^{*}$ events, where we use the same decay modes listed in
Table~\ref{tab:Ntags}; we find these events and then examine the
signal shape in data when one $D_s$ is ignored. The MM$^{*2}$
distribution from this sample is shown in
Fig.~\ref{data-Ds-MM2-DoubleTags}. The signal is fit to a Crystal
Ball function \cite{CBL,taunu}. The $\sigma$ parameter, that
represents the width of the distribution, is found to be
0.032$\pm$0.002 GeV$^2$. We do expect this to vary somewhat
depending on the final state, but we do not expect the parameters
that fix the shape of the tail to change, since they depend mostly
on beam radiation and the properties of photon detection.

We fit the MM$^{*2}$ distributions for each mode using the Crystal
Ball function with fixed tail parameters, but allowing $\sigma$ to
float, and a 5th order Chebyshev polynomial background. We find a
total of 18645$\pm$426 events within a $\pm 2.5\sigma$ interval
defined by the fit to each mode. There is also a small enhancement
of 4.8\% on our ability to find tags in $\mu^+\nu$ (or $\tau^+\nu$,
$\tau^+\to\pi^+\overline{\nu}$) events (tag bias) as compared with
generic events, determined by Monte Carlo simulation, to which we
assign a systematic error of 21\% giving a correction of $(4.8 \pm
1.0)$\%. An overall systematic error of 5\% on the number of tags is
assigned by changing the fitting range, using 4th order and 6th
order Chebychev background polynomials, and allowing the parameters
of the tail of the fitting function to float.

\subsection{Signal Reconstruction}
We next describe the search for $D_s^+\to\mu^+\nu$. Candidate events
are selected that contain only a single extra track with opposite
sign of charge to the tag. The track must make an angle
$>$35.9$^{\circ}$ with respect to the beam line, and in addition we
require that there not be any neutral cluster detected in the
calorimeter with energy greater than 300 MeV. These cuts are highly
effective in reducing backgrounds. The photon energy cut is
especially useful to reject $D_s^+\to \pi^+\pi^0$, should this mode
be significant, and $D_s^+\to\eta\pi^+$.

Since we are searching for events where there is a single missing
neutrino, the missing mass squared, MM$^2$, evaluated by taking into
account the observed $\mu^+$, $D_s^-$, and $\gamma$ should peak at
zero; the MM$^2$ is computed as

\begin{equation}
\label{eq:mm2} {\rm MM}^2=\left(E_{\rm
CM}-E_{D_s}-E_{\gamma}-E_{\mu}\right)^2
           -\left(\overrightarrow{p}_{\!\rm CM}-\overrightarrow{p}_{\!D_s}
           -\overrightarrow{p}_{\!\gamma}
           -\overrightarrow{p}_{\!\mu}\right)^2,
\end{equation}
where $E_{\mu}$ ($\overrightarrow{p}_{\!\mu}$) are the energy
(momentum) of the candidate muon track and all variables are the
same as defined in Eq.~\ref{eq:mmss}.

We also make use of a set of kinematical constraints and fit each
event to two hypotheses one of which is that the $D_s^-$ tag is the
daughter of a $D_s^{*-}$ and the other that the $D_s^{*+}$ decays
into $\gamma D_s^+$, with the $D_s^+$ subsequently decaying into
$\mu^+\nu$. The kinematical constraints, in the center-of-mass
frame, are
\begin{eqnarray}
\label{eq:constr}
&&\overrightarrow{p}_{\!D_s}+\overrightarrow{p}_{\!D_s^*}=0
\\\nonumber &&E_{\rm CM}=E_{D_s}+E_{D_s^*}\\\nonumber
&&E_{D_s^*}=\frac{E_{\rm
CM}}{2}+\frac{M_{D_s^*}^2-M_{D_s}^2}{2E_{\rm CM}}{\rm~or~}
E_{D_s}=\frac{E_{\rm CM}}{2}-\frac{M_{D_s^*}^2-M_{D_s}^2}{2E_{\rm
CM}}\\\nonumber &&M_{D_s^*}-M_{D_s}=143.6 {\rm ~MeV}.
\end{eqnarray}
In addition, we constrain the invariant mass of the $D_s^-$ tag to
the known $D_s$ mass. This gives us a total of 7 constraints. The
missing neutrino four-vector needs to be determined, so we are left
with a three-constraint fit. We perform a standard iterative fit
minimizing $\chi^2$. As we do not want to be subject to systematic
uncertainties that depend on understanding the absolute scale of the
errors, we do not make a $\chi^2$ cut but simply choose the photon
and the decay sequence in each event with the minimum $\chi^2$.

In this analysis, we consider three separate cases: (i) the track
deposits $<$~300 MeV in the calorimeter, characteristic of a
non-interacting pion or a muon; (ii) the track deposits $>$~300 MeV
in the calorimeter, characteristic of an interacting pion, and is
not consistent with being an electron; (iii) the track satisfies our
electron selection criteria defined below. Then we separately study
the MM$^2$ distributions for these three cases. The separation
between muons and pions is not complete. Case (i) contains 99\% of
the muons but also 60\% of the pions, while case (ii) includes 1\%
of the muons and 40\% of the pions \cite{DptomunPRD}. Case (iii)
does not include any signal but is used later for background
estimation. For cases (i) and (ii) we insist that the track not be
identified as a kaon. For electron identification we require a match
between the momentum measurement in the tracking system and the
energy deposited in the CsI calorimeter and we also require that
$dE/dx$ and RICH information be consistent with expectations for an
electron.

\subsection{The Expected MM$^2$ Spectrum}
\begin{figure}[hbt]
\centering
\includegraphics[width=5in]{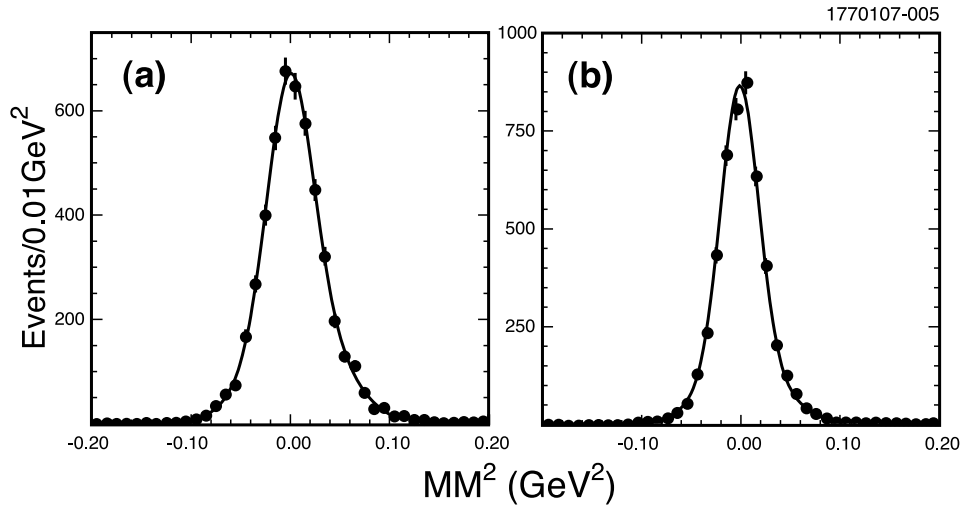}
\caption{The MM$^2$ resolution from Monte Carlo simulation for
$D_s^+\to\mu^+\nu$ utilizing a $\phi\pi^-$ tag and a $\gamma$ from
either $D_s^*$ decay, both before the kinematic fit (a) and after
(b).} \label{mc-munu-res}
\end{figure}

For the $\mu^+\nu$ final state the MM$^2$ distribution can be
modeled as the sum of two Gaussians centered at zero (see
Eq.~\ref{eq:twoGauss}). A Monte Carlo simulation of the MM$^2$ for
the $\phi\pi^-$ subset of $K^+K^-\pi^-$ tags is shown in
Fig.~\ref{mc-munu-res} both before and after the fit. The fit
changes the resolution from $\sigma$=0.032 GeV$^2$ to $\sigma$=0.025
GeV$^2$, a 22\% improvement, without any loss of events.

\begin{figure}[hbt]
\centering
\includegraphics[width=4in]{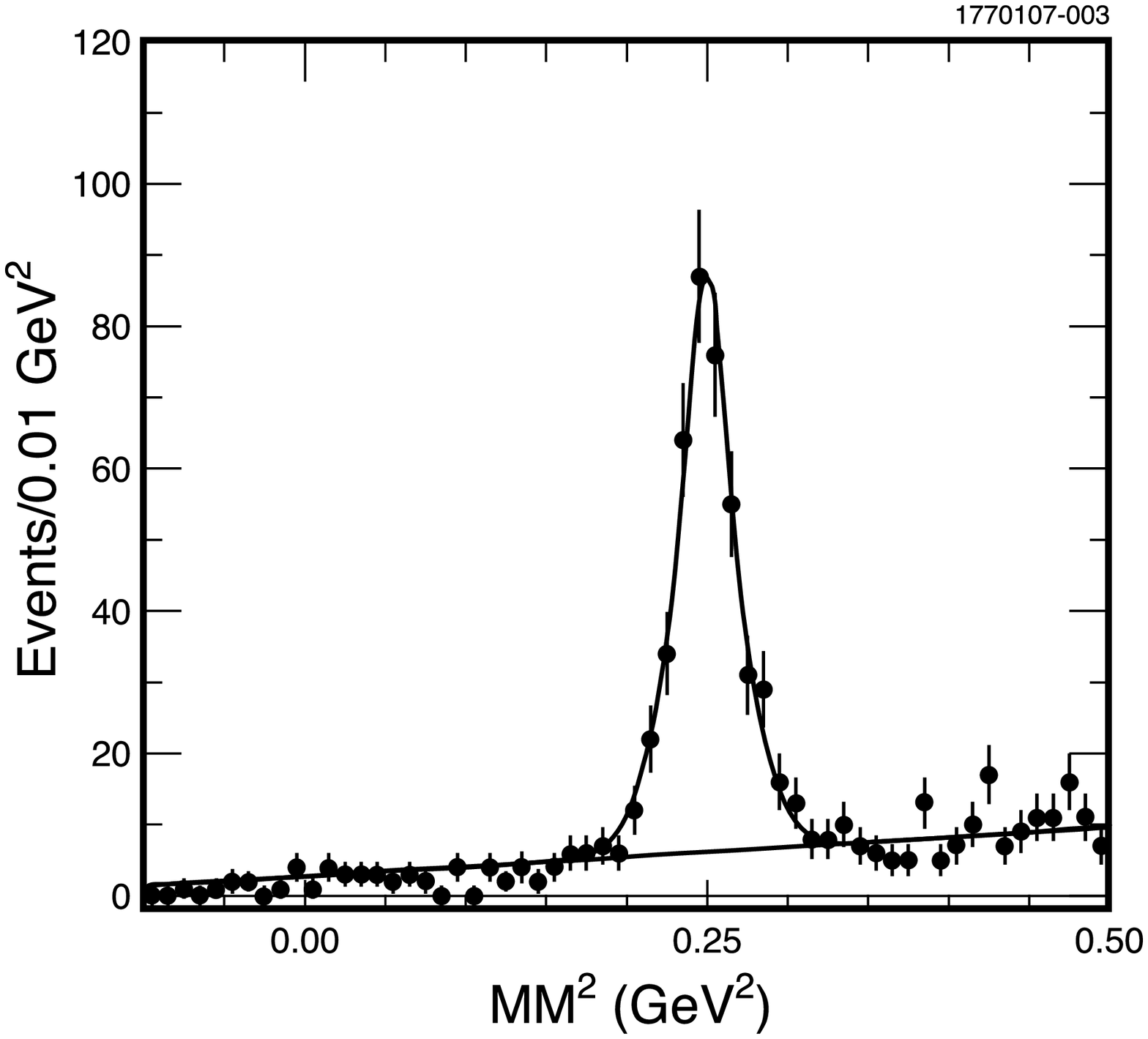}
\caption{The MM$^2$ distribution for events with an identified $K^+$
track. The kinematic fit has been applied. The curve is a fit to the
sum of two Gaussians centered at the square of the ${K}^0$ mass and
a linear background.} \label{Kmm2-mc-data}
\end{figure}

We check the resolution using data. The mode $D_s^+\to
\overline{K}^0K^+$ provides an excellent testing ground.\footnote{In
this paper the notation $\overline{K}^0K^+$ refers to the sum of
$\overline{K^0}K^+$ and ${K^0}K^+$ final states.} We search for
events with at least one additional track identified as a kaon using
the RICH detector, in addition to a $D_s^-$ tag. The MM$^2$
distribution is shown in Fig.~\ref{Kmm2-mc-data}. Fitting this
distribution to a two-Gaussian shape gives a MM$^2$ resolution of
0.025 GeV$^2$ in agreement with Monte Carlo simulation.

\begin{figure}[hbt]
\centering
\includegraphics[width=3in]{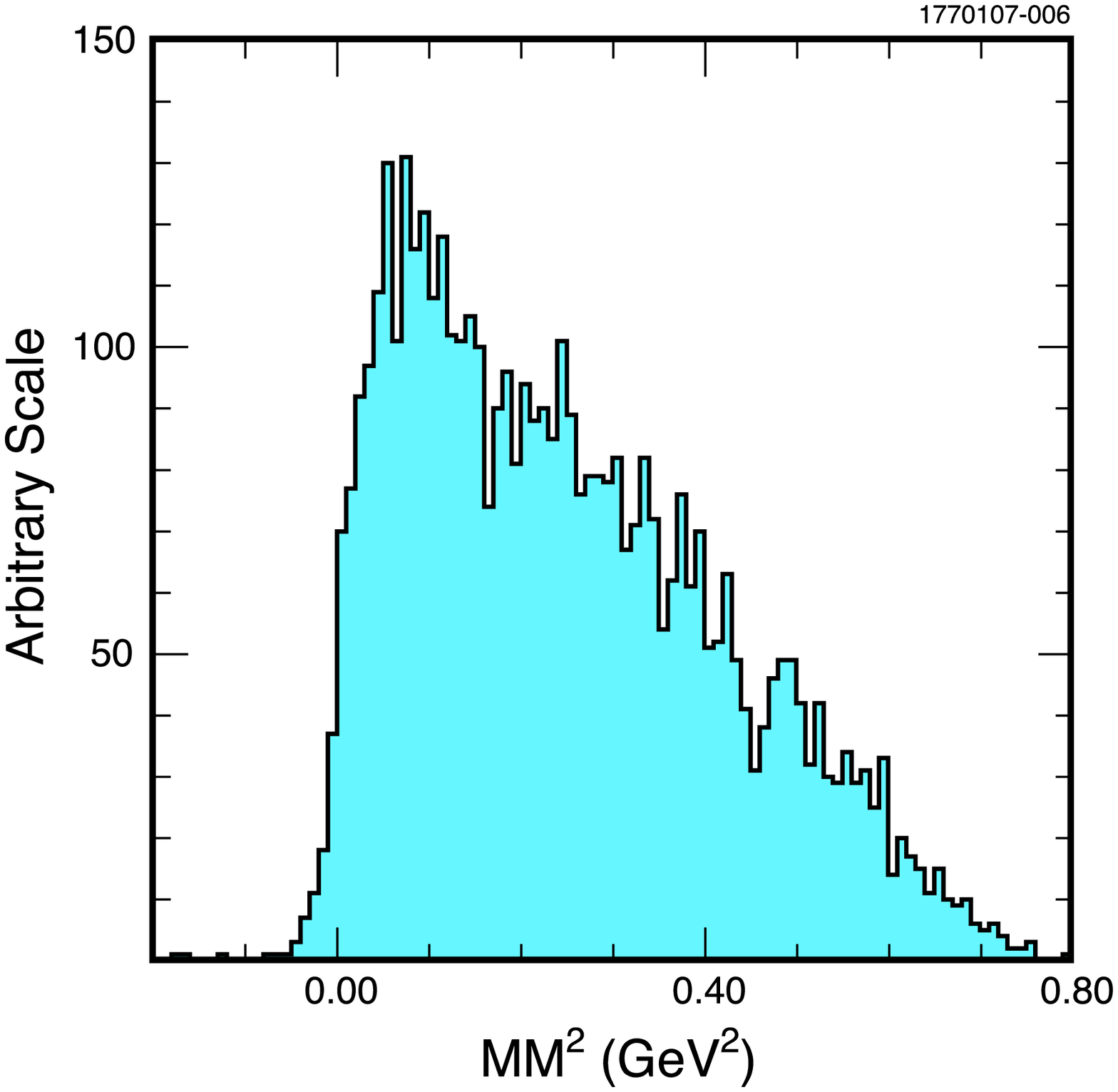}
\caption{The MM$^2$ distribution from Monte-Carlo simulation for
$D_s^+\to\tau^+\nu$, $\tau^+\to\pi^+\overline{\nu}$ at an $E_{\rm
CM}$ of 4170 MeV.} \label{mm2-taunu-pinu-mc}
\end{figure}

For the $\tau^+\nu$, $\tau^+\to\pi^+\overline{\nu}$ final state a
Monte Carlo simulation of the MM$^2$ spectrum is shown in
Fig.~\ref{mm2-taunu-pinu-mc}. The extra missing neutrino results in
a smeared distribution.

\subsection{MM$^2$ Spectra in Data}

\begin{figure}[htb]
\centerline{ \epsfxsize=3.5in \epsffile{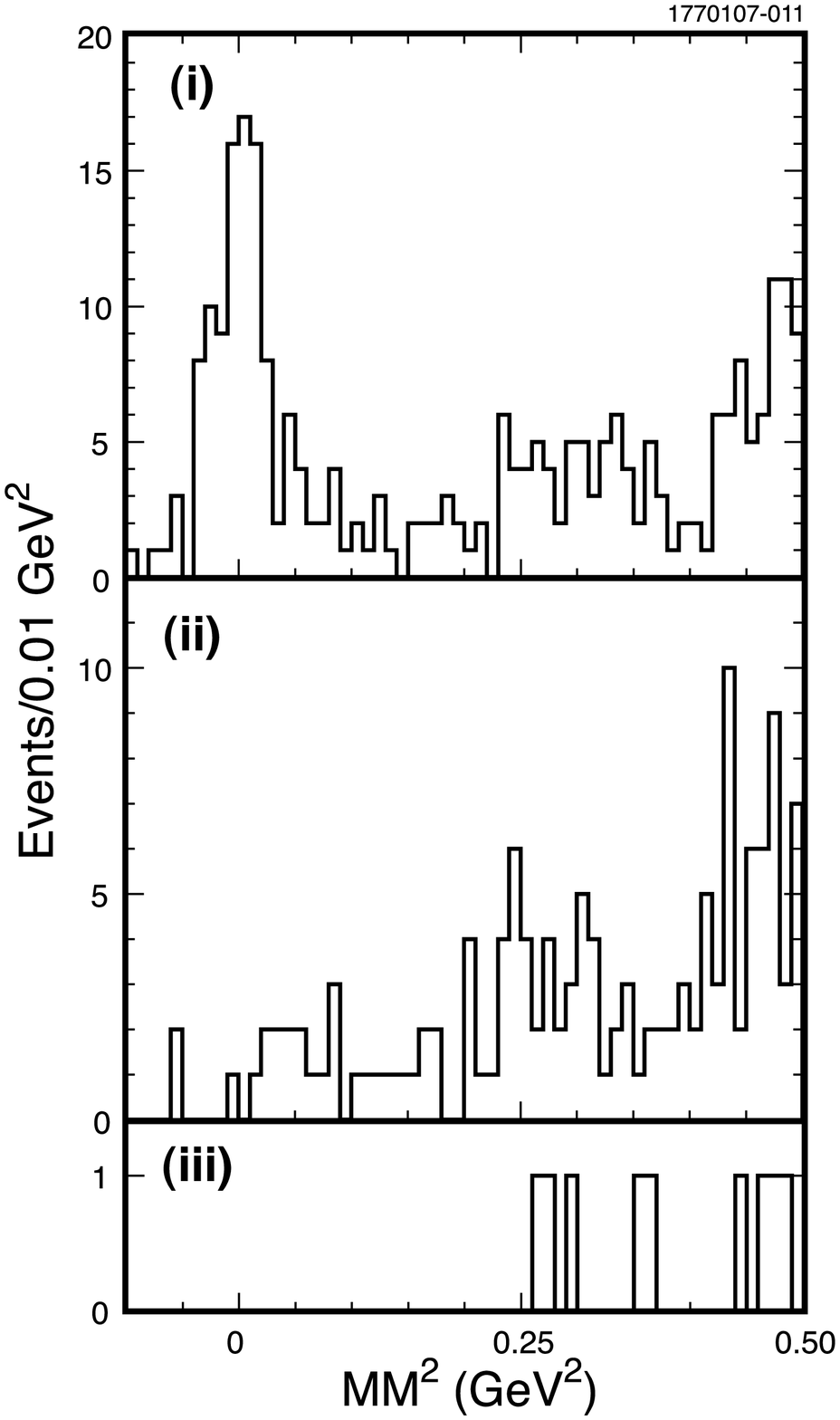}}
 \caption{The MM$^2$ distributions from data for events with a
$D_s^-$ reconstructed in a tag mode, an additional positively
charged track and no neutral energy clusters above 300 MeV.
 For case
(i) when the single track deposits $<$~300 MeV of energy in the
calorimeter. The peak near zero is from $D_s^+\to\mu^+\nu$ events.
Case (ii): track deposits $>$~300 MeV in the crystal calorimeter but
is not consistent with being an electron. Case (iii): the track is
identified as an electron. } \label{mm2-data}
 \end{figure}
 The MM$^2$ distributions from data are shown in Fig.~\ref{mm2-data}.
The overall signal region we consider is $-0.05<$ MM$^2$ $<$ 0.20
GeV$^2$. The higher limit is imposed to exclude background from
$\eta\pi^+$ and ${K}^0\pi^+$ final states. There is a clear peak in
Fig.~\ref{mm2-data}(i) due to $D_s^+\to\mu^+\nu$. Furthermore, the
region between the $\mu^+\nu$ peak and 0.20 GeV$^2$ has events that
we will show are dominantly due to the $D_s^+\to\tau^+\nu$ decay.
The events in Fig.~\ref{mm2-data}(ii) below 0.20 GeV$^2$ are also
dominantly due to $\tau^+\nu$ decay.

The specific signal regions are defined as follows: for $\mu^+\nu$,
$-0.05<$MM$^2<0.05$ GeV$^2$, corresponding to $\pm 2\sigma$; for
$\tau\nu$, $\tau^+\to\pi^+\overline{\nu}$, in case (i)
$0.05<$MM$^2<0.20$ GeV$^2$ and in case (ii) $-0.05<$MM$^2<0.20$
GeV$^2$. In these regions we find 92, 31, and 25 events,
respectively.

\subsection{Background Evaluations}
\label{sec:background}

We consider the background arising from two sources: one from real
$D_s^+$ decays and the other from the background under the
single-tag signal peaks. For the latter, we obtain the background
from data.
 We define side-bands of the invariant mass signals shown in
Fig.~\ref{mbc} starting at 4$\sigma$ on the low and high sides of
the invariant mass peaks for all modes. The intervals extend away
from the signal peaks by approximately the same width used in
selecting the signal, 5$\sigma$, so as to ensure that the number
of background events in the sidebands accurately reflects the
numbers under the signal peaks. Thus the amount of data
corresponds to twice the number of background events under the
signal peaks, except for the $\eta\pi^-$ and $\eta\rho^-$ modes,
where the signal widths are so wide that we chose narrower
side-bands only equaling the data.  We analyze these events in
exactly the same manner as those in the signal peak.\footnote{The
$D_s$ mass used in the fit is chosen to be the middle of the
relevant sideband interval.}

The backgrounds are given here as the sum of two numbers, the first
being the number from all modes, except $\eta\pi^-$ and
$\eta\rho^-$, and the second being the number from these modes. 
For case (i) we find 2.5+1 background in the $\mu^+\nu$ signal
region and 2.5+0 background in the $\tau^+\nu$ region. For case (ii)
we find 2+1 events. Our total sideband background summing over all
of these cases is 9.0$\pm$2.3. The numbers of signal and background
events due to false $D_s^-$ tags as evaluated from sidebands are
given in Table~\ref{tab:signbk}.

\begin{table}[htb]
\begin{center}
\caption{Numbers of events in the signal region, and background
events evaluated from sideband regions. \label{tab:signbk}}
\begin{tabular}{cccc}
 \hline\hline
Case & Region (GeV$^2$)  & Signal & Background \\\hline
i & -0.05$<$MM$^2<0.05$  &  92  &  3.5$\pm$1.4\\
i &  0.05$<$MM$^2<0.20$  &  31  &  2.5$\pm$1.1\\
ii & -0.05$<$MM$^2<0.20$  & 25  &  3.0$\pm$1.3\\\hline

Sum & -0.05$<$MM$^2<0.20$ & 148 & 9.0$\pm$2.3\\\hline\hline
\end{tabular}
\end{center}
\end{table}
 This entire procedure
was checked by performing the same study on a sample of Monte
Carlo generated at an $E_{\rm CM}$ of 4170 MeV that includes known
charm and continuum production cross-sections. The Monte Carlo
sample corresponds to an integrated luminosity that is four times
larger than the data. We find the number of background events
predicted directly by examining the decay generator of the
simulation is 28 and the sideband method yields 22. These are
slightly smaller than found in the data, but consistent within
errors. We note that the Monte Carlo is far from perfect as many
branching fractions are unknown and so are estimated.

\begin{figure}[hbt]
\centering
\includegraphics[width=3in]{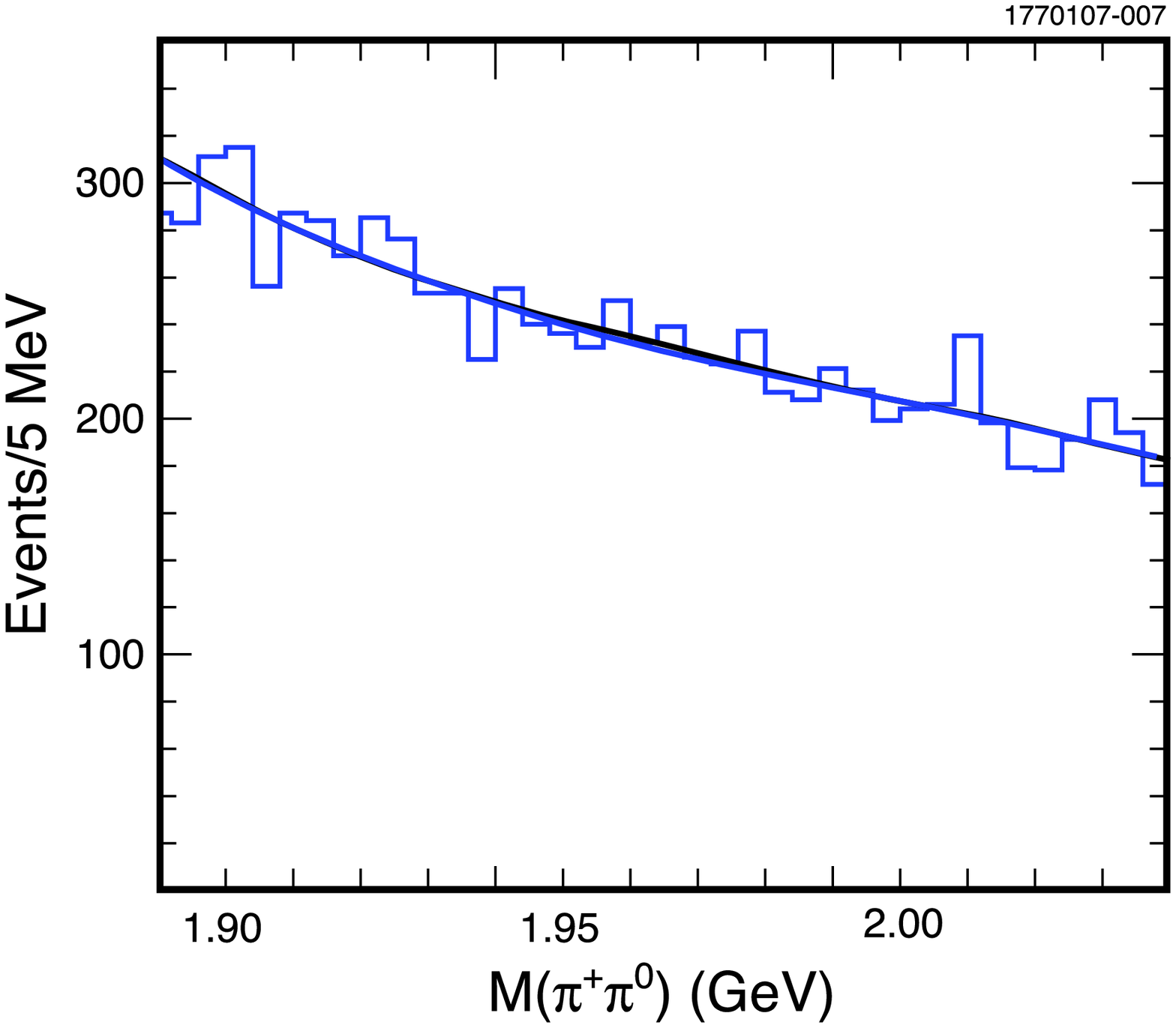}
\caption{The invariant $\pi^+\pi^0$ mass. The upper curve shows a
fit using a background polynomial plus Gaussian signal functions,
where the width of the Gaussian is fixed to a value determined by
Monte Carlo simulation. The lower curve shows just the background
polynomial. } \label{pipi0-single-tag}
\end{figure}

The background from real $D_s^+$ decays is studied by identifying
each possible source mode by mode. For the $\mu^+\nu$ final state,
the only possible background within the signal region is
$D_s^+\to\pi^+\pi^0$. This mode has not been studied previously.
We show in Fig.~\ref{pipi0-single-tag} the $\pi^+\pi^0$ invariant
mass spectrum from a 195 pb$^{-1}$ subsample of our data. We do
not see a signal and set an upper limit $<1.1\times 10^{-3}$ at
90\% confidence level. Recall that any such events are also
heavily suppressed by the extra photon energy cut of 300 MeV.
There are also some $D_s^+\to\tau^+\nu$,
$\tau^+\to\pi^+\overline{\nu}$ events that occur in the signal
region. Using the SM expected ratio of decay rates from
Eq.~\ref{eq:tntomu} we calculate a contribution of 7.4
$\pi^+\overline{\nu}\nu$ events that we will treat as part of the
signal.

For the $\tau^+\nu$, $\tau^+\to\pi^+\overline{\nu}$ final state the
real $D_s^+$ backgrounds include, in addition to the $\pi^+\pi^0$
background discussed above, semileptonic decays, possible
$\pi^+\pi^0\pi^0$ decays, and other $\tau^+$ decays. Semileptonic
decays involving muons are equal to those involving electrons shown
in Fig.~\ref{mm2-data}(c). Since no electron events appear in the
signal region, the background from muons is taken to be zero. The
$\pi^+\pi^0\pi^0$ background is estimated by considering the
$\pi^+\pi^+\pi^-$ final state whose measured branching fraction is
(1.02$\pm$0.12)\% \cite{stone-fpcp}. This mode has large
contributions from $f_0(980)\pi^+$ and other $\pi^+\pi^-$ resonant
structures at higher mass \cite{focus-3pi}. The $\pi^+\pi^0\pi^0$
mode will also have these contributions, but the MM$^2$ opposite to
the $\pi^+$ will be at large mass. The only component that can
potentially cause background is the non-resonant component measured
by FOCUS as (17$\pm$4)\% \cite{focus-3pi}. This background has been
evaluated by Monte Carlo simulation as have backgrounds from other
$\tau^+$ decays, and each is listed in Table~\ref{tab:taunubkrd}.

\begin{table}[htb]
\begin{center}
\caption{Backgrounds in the $D_s^+\to\tau^+\nu$,
$\tau^+\to\pi^+\overline{\nu}$ sample for correctly reconstructed
tags, case (i) for 0.05$<$MM$^2<0.20$ GeV$^2$ and case (ii) for
-0.05$<$MM$^2<0.20$ GeV$^2$. \label{tab:taunubkrd}}
\begin{tabular}{lcccc}
 \hline\hline
    Source  & ${\cal{B}}$(\%)            &  \# of events case (i) &  \# of events case(ii) & Sum\\ \hline
$D_s^+\to X \mu^+\nu$  & 8.2  & 0$^{+1.8}_{-0}$ & 0 & 0$^{+1.8}_{-0}$\\
$D_s^+\to\pi^+\pi^0\pi^0 $ & 1.0  & 0.03$\pm$0.04 & 0.08$\pm$0.03 & 0.11$\pm$0.04\\
$D_s^+\to\tau^+\nu$ & 6.4 & & &\\
~~~~$\tau^+\to \pi^+\pi^0\overline{\nu}$ & 1.5 & 0.55$\pm$0.22 & 0.64$\pm$0.24 & 1.20$\pm$0.33\\
~~~~$\tau^+\to \mu^+\overline{\nu}\nu$ & 1.0 & 0.37$\pm$0.15 & 0 &0.37$\pm$0.15 \\
\hline
Sum & & 1.0$^{+1.8}_{-0}$ & 0.7$\pm$0.2 & 1.7$^{+1.8}_{-0.4}$\\
\hline\hline
\end{tabular}
\end{center}
\end{table}

\section{Leptonic Branching Fractions}

\begin{figure}[htbp]
 \vskip 0.00cm
\centerline{ \epsfxsize=4.0in
\epsffile{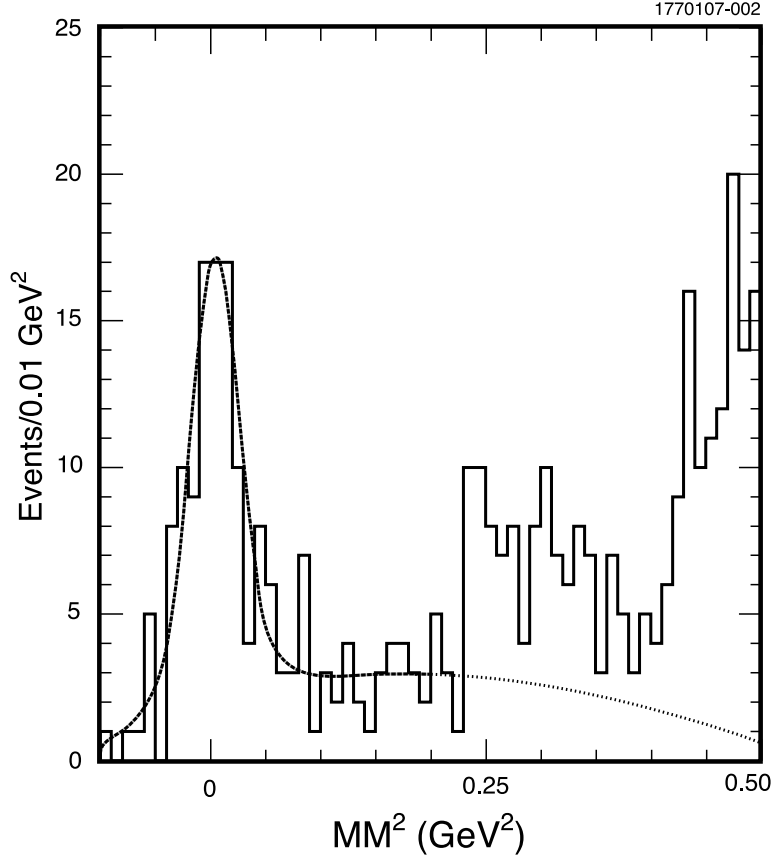} }
 \caption{The sum of case (i) and case (ii) MM$^2$ distributions (histogram) compared to
 the predicted shape (curve) for the sum of $D_s^+\to \mu^+\nu$ and $\tau^+\nu$,
 $\tau^+\to \pi^+\overline{\nu}$. The curve is
normalized to the total number of events below MM$^2<$0.2 GeV$^2$.
} \label{try1-total}
 \end{figure}

The sum of MM$^2$ distributions for case (i) and case (ii),
corresponding to the sum of $D_s^+\to \mu^+\nu$ and $D_s^+\to
\tau^+\nu$, $\tau^+\to \pi^+\overline{\nu}$ candidates, is compared
in Fig.~\ref{try1-total} with the expected shape, assuming the SM
value of $R$ as given in Eq.~\ref{eq:tntomu} for the ratio of
$\tau^+\nu$ to $\mu^+\nu$ rates. The curve is normalized to the
total number of events below MM$^2<$0.2 GeV$^2$. Besides the
prominent $\mu^+\nu$ peak and $\tau^+\nu$;
$\tau^+\to\pi^+\overline{\nu}$ shoulder, there is an enhancement
between 0.25-0.35 GeV$^2$, due to $K^0\pi^+$ and $\eta\pi^+$ final
states, where the decay products other than the $\pi^+$ escape
detection. The data are consistent with our expectation that the
region $-0.05<$MM$^2<$0.2 GeV$^2$ contains mostly signal. Recall
there are 148 total events only 10.7 of which we estimate are
background, 9.0 from fake $D_s^-$ tags and 1.7 from real tags and
$D_s^+$ decays. Above 0.2 GeV$^2$ other, larger backgrounds enter.

The number of real $\mu^+\nu$ events $N_{\mu\nu}$ is related to the
number of events detected in the signal region $N_{\rm det}$ (92),
the estimated background $N_{\rm bkgrd}$ (3.5), the number of tags,
$N_{\rm tag}$, and the branching fractions as
\begin{equation}
 N_{\mu\nu}\equiv N_{\rm det}-N_{\rm bkgrd}=N_{\rm tag}\cdot
\epsilon\left[\epsilon'{\cal{B}}(D_s^+\to
\mu^+\nu)+\epsilon''{\cal{B}}(D_s^+\to
\tau^+\nu;~\tau^+\to\pi^+\overline{\nu})\right], \label{eq:munuB}
\end{equation}
where $\epsilon$ (80.1\%) includes the efficiency for reconstructing
the single charged track including final state radiation (77.8\%),
the (98.3$\pm$0.2)\% efficiency of not having another unmatched
cluster in the event with energy greater than 300 MeV, and for the
fact that it is easier to find tags in $\mu^+\nu$ events than in
generic decays by 4.8\%, as determined by Monte Carlo simulation.
The efficiency labeled $\epsilon'$ (91.4\%) is the product of the
99.0\% muon efficiency for depositing less than 300 MeV in the
calorimeter and 92.3\% acceptance of the MM$^2$ cut of $|$MM$^2|<
0.05$ GeV$^2$. The quantity $\epsilon''$ (7.9\%) is the fraction of
$\tau^+\nu;~\tau^+\to\pi^+\overline{\nu}$ events contained in the
$\mu^+\nu$ signal window (13.2\%) times the 60\% acceptance for a
pion to deposit less than 300 MeV in the calorimeter.

The two $D_s^+$ branching fractions in Eq.~\ref{eq:munuB} are
related as
\begin{equation}
{\cal{B}}(D_s^+\to
\tau^+\nu;~\tau^+\to\pi^+\overline{\nu})=R\cdot{\cal{B}}(\tau^+\to\pi^+\overline{\nu}){\cal{B}}(D_s^+\to
\mu^+\nu)=1.059\cdot{\cal{B}}(D_s^+\to \mu^+\nu)~,
\end{equation}
where we take the Standard Model ratio for $R$ as given in
Eq.~\ref{eq:tntomu} and
${\cal{B}}(\tau^+\to\pi^+\overline{\nu})$=(10.90$\pm$0.07)\%
\cite{PDG}. This allows us to solve Eq.~\ref{eq:munuB}. Since
$N_{\rm det}$= 92, $N_{\rm bkgrd}$=3.5$\pm$1.4, and $N_{\rm
tag}=18645\pm 426\pm 1081$, we find
\begin{equation}
{\cal{B}}(D_s^+\to \mu^+\nu)= (0.594\pm 0.066\pm0.031)\%.
\end{equation}

We can also sum the $\mu^+\nu$ and $\tau^+\nu$ contributions, where
we restrict ourselves to the MM$^2$ region below 0.20 GeV$^2$ and
above -0.05 GeV$^2$. Eq.~\ref{eq:munuB} still applies. The number of
events in the signal region and the number of background events
changes to 148 and 10.7$^{+2.9}_{-2.3}$, respectively. The
efficiency $\epsilon'$ becomes 96.2\%, and $\epsilon''$ increases to
45.2\%. Using this method, we find an effective branching fraction
of
\begin{equation}
{\cal{B}}^{\rm eff}(D_s^+\to \mu^+\nu)=(0.638 \pm 0.059 \pm 0.033)\%
. \label{eq:finalBR}
\end{equation}

The systematic errors on these branching fractions are given in
Table~\ref{tab:munusys}. The error on track finding is determined
from a detailed comparison of the simulation with double tag
events where one track is ignored. ``Minimum ionization" indicates
the error due to the requirement that the charged track deposit no
more than 300 MeV in the calorimeter; it is determined using
two-body $D^0\to K^-\pi^+$ decays (see Ref.~\cite{DptomunPRD}).
The error on the photon veto efficiency, due to the 300 MeV/c
extra shower energy cut, is determined using Monte Carlo
simulation. The Monte Carlo was cross-checked using a sample of
fully reconstructed $D^+D^-$ events and comparing the inefficiency
due to additional photons with energy above 300 MeV/c.  These
events have no real extra photons above 300 MeV/c; those that are
present are due to interactions of the $D^{\pm}$ decay products in
the detector material. The error on the number of tags of $\pm$5\%
has been discussed earlier. In addition there is a small error of
$\pm$0.6\% on the $\tau^+\nu$ branching fraction due to the
uncertainty on the $\tau^+$ decay fraction to
$\pi^+\overline{\nu}$. Additional systematic errors arising from
the background estimates are negligible. Note that the minimum
ionization error does not apply to the summed branching fraction
given in Eq.~\ref{eq:finalBR}; in this case the total systematic
error is 5.1\%.
\begin{table}[htb]
\begin{center}
\caption{Systematic errors on determination of the $D_s^+\to
\mu^+\nu$ branching fraction. \label{tab:munusys}}
\begin{tabular}{lc} \hline\hline
   Error Source & Size (\%) \\ \hline
Track finding &0.7 \\
Photon veto & 1 \\
Minimum ionization$^*$ & 1\\
Number of tags& 5\\
\hline
Total & 5.2\\
 \hline\hline
\end{tabular}
\end{center}
*-Not applicable for summed rate
\end{table}

We also analyze the $\tau^+\nu$ final state independently. We use
different MM$^2$ regions for cases (i) and (ii) defined above. For
case (i) we define the signal region to be the interval 0.05$<$MM$^2
<$0.20 GeV$^2$, while for case (ii) we define the signal region to
be the interval
 -0.05$<$MM$^2<$0.20 GeV$^2$. Case (i) includes  the $\mu^+\nu$ signal,
 so we must exclude the region close to zero MM$^2$, while for case (ii) we are specifically
 selecting pions so the signal region can be larger.
 The upper limit on MM$^2$ is
 chosen to avoid background from the tail of the ${K}^0\pi^+$ peak.
 The fractions of the MM$^2$
 range accepted are 32\% and 45\% for case (i) and (ii), respectively.

 We find 31 signal and 3.5$^{+1.7}_{-1.1}$ background events for case (i) and 25 signal and
 5.1$\pm$1.6 background events
 for case (ii). The branching fraction, averaging the two cases is
\begin{equation}
{\cal{B}}(D_s^+\to \tau^+\nu)=(8.0\pm 1.3\pm0.4)\% .
\end{equation}
Lepton universality in the Standard Model requires that the ratio
$R$ from Eq.~\ref{eq:tntomu} be equal to a value of 9.72. We measure
\begin{equation}
R\equiv \frac{\Gamma(D_s^+\to \tau^+\nu)}{\Gamma(D_s^+\to
\mu^+\nu)}= 13.4\pm 2.6 \pm 0.2~. \label{eq:tntomu2}
\end{equation}
Here the systematic error is dominated by the uncertainty on the
minimum ionization cut that we use to separate the $\mu^+\nu$ and
$\tau^+\nu$ regions at 300 MeV. We take this error as 2\%, since a
change here affects both the numerator and denominator. The ratio is
consistent with the Standard Model prediction. Current results on
$D^+$ leptonic decays also show no deviations \cite{ourDptotaunu}.
The absence of any detected electrons opposite to our tags allows us
to set an upper limit of
\begin{equation}
{\cal{B}}(D_s^+\to e^+\nu)< 1.3\times 10^{-4}
\end{equation}
at 90\% confidence level; this is also consistent with Standard
Model predictions and lepton universality.

\section{Checks of the Method}

We perform an overall check of our procedures by measuring
${\cal{B}}(D_s^+\to \overline{K}^0K^+)$.  For this measurement we
compute the MM$^2$ (Eq.~\ref{eq:mm2}) using events with an
additional charged track but here identified as a kaon. These track
candidates have momenta of approximately 1 GeV/c; here our RICH
detector has a pion to kaon fake rate of 1.1\% with a kaon detection
efficiency of 88.5\% \cite{RICH}. For this study, we do not veto
events with extra charged tracks, or neutral energy deposits $>$300
MeV, because of the presence of the ${K^0}$.

The MM$^2$ distribution is shown in Fig.~\ref{Kmm2-mc-data}. The
peak near 0.25 GeV$^2$ is due to the decay mode of interest. We fit
this to a linear background from 0.02-0.50 GeV$^2$ plus a
two-Gaussian signal function. The fit yields 375$\pm$23$\pm$18
events. Events from the $\eta\pi^+$ mode where the $\pi^+$ fakes a
$K^+$ are very rare and would not peak at the proper MM$^2$.  Since
$\eta K^+$ could in principle contribute a background in this
region, we searched for this final state in a 195 pb$^{-1}$
subsample of the data. Not finding any signal, we set an upper limit
of ${\cal{B}}(D_s^+\to \eta K^+)< 2.8\times 10^{-3}$ at 90\%
confidence level, approximately a factor of ten below our
measurement. This final state would peak at a MM$^2$ of 0.30 GeV$^2$
and would cause an asymmetric tail on the high side of the peak.
Since we see no evidence for an asymmetry in the $\overline{K}^0K^+$
peak we ignore the $\eta K^+$ final state from here on. In order to
compute the branching fraction we must include the efficiency of
detecting the kaon track 76.2\%, including radiation
\cite{gammamunu}, the particle identification efficiency of 88.5\%,
and take into account that it is easier to detect tags in events
containing a $\overline{K}^0K^+$ decay than in the average
$D_sD_s^*$ event due to the track and photon multiplicities, which
gives a 3\% correction.\footnote{The tag bias is less here than in
the $\mu^+\nu$ case because of the $K^0$ decays and interactions in
the detector.} These rates are estimated by using Monte Carlo
simulation. We determine
\begin{equation}
\label{eq:KK0}
 {\cal{B}}(D_s^+\to \overline{K}^0K^+)=(2.90\pm0.19\pm
0.18)\%,
\end{equation}
where the systematic errors are listed in Table~\ref{tab:sysKK}.
We estimate the error from the signal shape by taking the change
in the number of events when varying the signal width of the
two-Gaussian function by $\pm 1\sigma$. The error on the
background shapes is given by varying the shape of the background
fit. The error on the particle identification efficiency is
measured using two-body $D^0$ decays \cite{RICH}. The other errors
are the same as described in Table~\ref{tab:munusys}. Again, the
largest component of the systematic error arises from the number
of tag events (5\%). In fact, to use this result as a check on our
procedures, we need only consider the systematic errors that are
different here than in the $\mu^+\nu$ case. Those are due only to
the signal and background shapes and the particle identification
cut. Those systematic errors amount to 3.7\% or $\pm$0.11 in the
branching fraction.

To determine absolute branching fractions of charm mesons, CLEO-c
uses a method where both particles are fully reconstructed (so
called ``double tags") and the rates are normalized using events
where only one particle is fully reconstructed. Our preliminary
result using this method for ${\cal{B}}(D_s^+\to K_SK^+)$=(1.50$\pm
0.09\pm 0.05)$\%, which when doubled becomes (3.00$\pm 0.19\pm
0.10)$\% \cite{Peter}. This is in excellent agreement with the
number in Eq.~\ref{eq:KK0}.
 These results are not independent.

\begin{table}[htb]
\begin{center}

\caption{Systematic errors on determination of the $D_s^+\to
\overline{K}^0K^+$ branching fraction. \label{tab:sysKK}}
\begin{tabular}{lc} \hline\hline
   Error Source & Size (\%) \\ \hline
Signal shape & 3\\
Background shape & 2\\
Track finding &0.7 \\
PID cut &1.0 \\
Number of tags& 5\\
\hline
Total & 6.3 \\
 \hline\hline
\end{tabular}
\end{center}
\end{table}

We also performed the entire analysis on a Monte Carlo sample that
corresponds to an integrated luminosity four 4 times larger than
the data sample. The input branching fraction in the Monte Carlo
is 0.5\% for $\mu^+\nu$ and 6.57\% for $\tau^+\nu$, while our
analysis measured
${\cal{B}}(D_s^+\to\mu^+\nu)=$(0.514$\pm$0.027)\% for the case (i)
$\mu^+\nu$ signal and (0.521$\pm$0.024)\% for $\mu^+\nu$ and
$\tau^+\nu$ combined. We also find (6.6$\pm$0.6)\% for the
$\tau^+\nu$ rate.

\section{The Decay Constant}

 Using our most precise value for ${\cal{B}}(D_s^+\to \mu^+\nu)$
 from Eq.~\ref{eq:finalBR}, that is derived using both our $\mu^+\nu$ and $\tau^+\nu$
 samples,
 and Eq.~\ref{eq:equ_rate} with a $D_s$ lifetime of (500$\pm$7)$\times 10^{-15}\,{\rm s}$ \cite{PDG}, we extract
 \begin{equation}
 f_{D_s^+}=274\pm 13 \pm 7 {~\rm MeV}.
 \end{equation}

We combine with our previous result \cite{our-fDp}
\begin{equation}
f_{D^+}=222.6\pm 16.7^{+2.8}_{-3.4}{\rm ~MeV}
\end{equation}
 and find a  value for
\begin{equation}
\displaystyle{\frac{f_{D_s^+}}{f_{D^+}}=1.23\pm 0.11\pm 0.04},
\end{equation}
where only a small part of the systematic error cancels in the ratio
of our two measurements.

\section{Conclusions}
Theoretical models that predict $f_{D_s^+}$ and the ratio
$\frac{f_{D_s^+}}{f_{D^+}}$ are listed in Table~\ref{tab:Models}.
Our result for $f_{D_s}$ is slightly higher than most theoretical
expectations. We are consistent with lattice gauge theory, and most
other models, for the ratio of decay constants. There is no evidence
at this level of precision for any suppression in the ratio due to
the presence of a virtual charged Higgs \cite{Akeroyd}.

\begin{table}[htb]
\begin{center}

\caption{Theoretical predictions of $f_{D^+_s}$, $f_{D^+}$, and
$f_{D_s^+}/f_{D^+}$. QL indicates quenched lattice calculations.}
\label{tab:Models}
\begin{tabular}{lccl} \hline\hline
    Model &$f_{D_s^+}$ (MeV) &  $f_{D^+}$ (MeV)          &  ~~~~~$f_{D_s^+}/f_{D^+}$           \\\hline
Lattice (HPQCD+UKQCD) \cite{Lat:Foll} & $241\pm 3$ & $208\pm 4$ &
$1.162\pm 0.009$\\
 Lattice (FNAL+MILC+HPQCD)  \cite{Lat:Milc} &
 $249 \pm 3 \pm 16 $&$201\pm 3 \pm 17 $&$1.24\pm 0.01\pm 0.07$ \\
QL (QCDSF) \cite{QCDSF} &
$220\pm 6 \pm 5 \pm 11$ &$206 \pm 6\pm 3\pm 22 $&$1.07\pm 0.02\pm 0.02$ \\
 QL (Taiwan) \cite{Lat:Taiwan} &
$266\pm 10 \pm 18$ &$235 \pm 8\pm 14 $&$1.13\pm 0.03\pm 0.05$ \\
QL (UKQCD) \cite{Lat:UKQCD}&$236\pm 8^{+17}_{-14}$ & $210\pm 10^{+17}_{-16}$ & $1.13\pm 0.02^{+0.04}_{-0.02}$\\
QL \cite{Lat:Damir} & $231\pm 12^{+6}_{-1}$&$211\pm 14^{+2}_{-12}$ &
$1.10\pm 0.02$\\
QCD Sum Rules \cite{Bordes} & $205\pm 22$ & $177\pm 21$ & $1.16\pm
0.01\pm 0.03$\\
QCD Sum Rules \cite{Chiral} & $235\pm 24$&$203\pm 20$ & $1.15\pm 0.04$ \\
Field Correlators \cite{Field} & $210\pm 10$&$260\pm 10$ & $1.24\pm 0.03$ \\
Quark Model \cite{Quarkmodel}&268 &$234$  & 1.15 \\
Quark Model \cite{QMII}&248$\pm$27 &$230\pm$25  & 1.08$\pm$0.01 \\
LFQM (Linear) \cite{Choi} & 211 & 248 & 1.18\\
LFQM (HO) \cite{Choi} &194 & 233  & 1.20\\
LF-QCD \cite{LF-QCD} & 253& 241  & 1.05 \\
Potential Model \cite{Equations} & 241& 238  & 1.01 \\
Isospin Splittings \cite{Isospin} & & $262\pm 29$ & \\
\hline\hline
\end{tabular}
\end{center}
\end{table}

By using a theoretical prediction for $f_{D_s^+}/f_{D^+}$ we can
derive a value for the ratio of CKM elements $|V_{cd}/V_{cs}|$.
Taking the value from Ref.~\cite{Lat:Foll} of $1.162\pm 0.009$, we
find
\begin{equation}
|V_{cd}/V_{cs}|=0.2171\pm 0.021\pm 0.0017~,
\end{equation}
where the first error is due the statistical and systematic errors
of the experiment and the second is due to the stated error on the
theoretical prediction. This value is expected to be almost equal to
the ratio of the CKM elements $|V_{us}|/|V_{ud}|$

We now compare with previous measurements. The branching fractions,
modes, and derived values of $f_{D_s^+}$ are listed in
Table~\ref{tab:fDs}. Our values are shown first. We are generally
consistent with previous measurements, although ours are more
precise.

\begin{table}[htb]
\begin{center}

\caption{Our results for ${\cal{B}}(D_s^+\to \mu^+\nu)$,
${\cal{B}}(D_s^+\to \tau^+\nu)$, and $f_{D_s^+}$ compared with
previous measurements. Results have been updated for the new value
of the $D_s$ lifetime \cite{PDG}. ALEPH combines both
measurements to derive a value for the decay
constant.\label{tab:fDs}}
\begin{tabular}{llccc}\hline\hline
Exp. & Mode  & ${\cal{B}}$& ${\cal{B}}_{\phi\pi}$ (\%) & $f_{D_s^+}$ (MeV) \\
\hline CLEO-c & $\mu^+\nu$ & $(5.94\pm 0.66\pm 0.31)\cdot 10^{-3}$ & & $264\pm 15\pm 7$\\
CLEO-c & $\tau^+\nu$ & $(8.0\pm 1.3\pm 0.4)\cdot 10^{-2}$&& $310\pm 25 \pm 8 $ \\
CLEO-c & combined & -& & $274\pm 13\pm 7$  \\
CLEO \cite{CLEO}& $\mu^+\nu$ &$(6.2\pm 0.8\pm 1.3 \pm 1.6)\cdot
10^{-3}$&
3.6$\pm$0.9&$273\pm19\pm27\pm33$\\
BEATRICE \cite{BEAT} & $\mu^+\nu$ &$(8.3\pm 2.3\pm 0.6 \pm
2.1)\cdot 10^{-3}$& 3.6$\pm$0.9&$312\pm43\pm12 \pm39$\\
ALEPH \cite{ALEPH}& $\mu^+\nu$ &$(6.8\pm 1.1\pm 1.8)\cdot 10^{-3}$ & 3.6$\pm$0.9& $282\pm 19\pm 40$ \\
ALEPH \cite{ALEPH}& $\tau^+\nu$ &$(5.8\pm 0.8\pm 1.8)\cdot 10^{-2}$ & &  \\
L3 \cite{L3} &$\tau^+\nu$ & $(7.4\pm 2.8 \pm 1.6\pm 1.8)\cdot 10^{-2}$ & & $299\pm 57\pm 32 \pm 37$  \\
OPAL \cite{OPAL} & $\tau^+\nu$ & $(7.0\pm 2.1 \pm 2.0)\cdot 10^{-2}$ & & $283\pm 44\pm 41$  \\
BaBar \cite{Babar-munu} & $\mu^+\nu$& $(6.74\pm 0.83\pm 0.26 \pm
0.66)\cdot 10^{-3}$ & 4.71$\pm$0.46 &
 $283\pm 17 \pm 7 \pm 14$\\\hline\hline
\end{tabular}
\end{center}

\end{table}

 Most measurements
of $D_s^+\to\ell^+\nu$ are normalized with respect to ${\cal{B}}
(D_s^+\to\phi\pi^+)\equiv {\cal{B}}_{\phi\pi}$.
 An exception is the OPAL measurement which is
normalized to the $D_s$ fraction in $Z^0$ events that is derived
from an overall fit to heavy flavor data at LEP \cite{HFAG}. It
still, however, relies on absolute branching fractions that are
hidden by this procedure, and the estimated error on the
normalization is somewhat smaller than that indicated by the error
on ${\cal{B}}_{\phi\pi}$ available at the time of their
publication. The L3 measurement is normalized taking the fraction
of $D_s$ mesons produced in $c$ quark fragmentation as
0.11$\pm$0.02, and the ratio of $D_s^*/D_s$ production of
0.65$\pm$0.10. The ALEPH results use ${\cal{B}}_{\phi\pi}$ for
their $\mu^+\nu$ results and a similar procedure as OPAL for their
$\tau^+\nu$ results. We note that the recent BaBar result uses a
larger ${\cal{B}}_{\phi\pi}$ than the other results.

The CLEO-c determination of $f_{D_s^+}$ is the most accurate to
date. It also does not rely on the independent determination of any
normalization mode. (We note that a preliminary CLEO-c result using
$D_s^+\to\tau^+\nu$, $\tau^+\to e^+\overline{\nu}\nu$ \cite{Moscow}
is consistent with these results.)

\section{Acknowledgments}

We gratefully acknowledge the effort of the CESR staff in providing
us with excellent luminosity and running conditions.
D.~Cronin-Hennessy and A.~Ryd thank the A.P.~Sloan Foundation. This
work was supported by the National Science Foundation, the U.S.
Department of Energy, and the Natural Sciences and Engineering
Research Council of Canada.

\end{document}